\documentclass[letter]{aa}

\usepackage{graphicx}
\usepackage{txfonts}
\usepackage{color}

\begin{document}

\title{Formation of a barred galaxy in a major merger:\\ The role of AGN feedback
}

\author{Ewa L. {\L}okas
}

\institute{Nicolaus Copernicus Astronomical Center, Polish Academy of Sciences,
Bartycka 18, 00-716 Warsaw, Poland\\
\email{lokas@camk.edu.pl}\\
 \\ }

\titlerunning{Formation of a barred galaxy in a major merger}
\authorrunning{E. L. {\L}okas}


\abstract{
Among the many processes involved in galaxy evolution, those of bar formation, quenching, and feedback from an active
galactic nucleus (AGN) seem to be connected, however, the nature of these relations remains unclear. In this
work, we aim to elucidate them by studying the formation of a barred galaxy in a major merger of two disks
in the IllustrisTNG simulations. This merger involves a coalescence of two supermassive black holes and a
sudden switch to the kinetic mode of AGN feedback implemented in the simulations, which leads to the removal of the gas
from the inner part of the galaxy, followed by quenching of star formation and the formation of the bar. This
causal relation between AGN feedback and bar formation explains a number of correlations observed in the data, such as
the higher frequency of bars among red spirals and the presence of central gas holes in barred galaxies. In such
a picture, the bars do not feed the black holes, so their presence does not increase the AGN strength, and
they do not cause quenching. However, bars do form in regions characterized by a low gas fraction resulting from AGN
feedback. This scenario is probably applicable to many barred galaxies, not only those formed in major mergers.}


\maketitle

\section{Introduction}

The relation between the processes of bar formation, quenching, and feedback from an active galactic nucleus (AGN) in
galaxy evolution remains an open question. It was recently established via cosmological simulations that AGN feedback
plays an important role in stopping star formation in galaxies mostly by removing gas from their inner parts
\citep{Bower2006, Dubois2015, Trayford2016, Weinberger2018, Dave2019, Xu2022, Piotrowska2022, Lokas2022b}. On the other
hand, simulations have also shown that the removal of the gas favors the formation of a bar in a galaxy
\citep{Athanassoula2013, Lokas2020a}. Indeed, observations of red disks have shown that they are more likely to possess
such a component \citep{Masters2011}. It is not clear, however, whether bars form preferentially in disks devoid of gas
or whether the bars form first, then moving the gas and leading to star formation quenching \citep{Newnham2020}. It has
also been suggested that supermassive black holes (SMBHs) may be fed by bars that trigger the motion of the gas toward
the center of the galaxy \citep{Emsellem2015}. Observations have demonstrated, rather, that the presence of a bar has
no significant influence on the AGN strength on the scale of the local Universe or at higher redshifts
\citep{Cisternas2015, Galloway2015, Cheung2015}.

In this letter, we attempt to elucidate the connection between these processes using a simulated galaxy from the
IllustrisTNG project \citep{Springel2018, Marinacci2018, Naiman2018, Nelson2018, Pillepich2018}. The simulations follow
the evolution of galaxies from early time to the present by solving for the gravity and magnetohydrodynamics, and
applying subgrid physics in the form of prescriptions for star formation, galactic winds, magnetic fields, and AGN
feedback. The applied model for AGN feedback includes the thermal mode dominating at high accretion rates and the
kinetic mode at low accretion rates, with the latter turned on when the black hole mass exceeds about $10^8$ M$_\odot$
\citep{Weinberger2017, Weinberger2018}. In the kinetic mode, feedback energy is injected into the surrounding gas in
the form of directed pulses and not continuously (as is the case in the thermal mode).

\begin{figure}
\centering
\includegraphics[width=7.3cm]{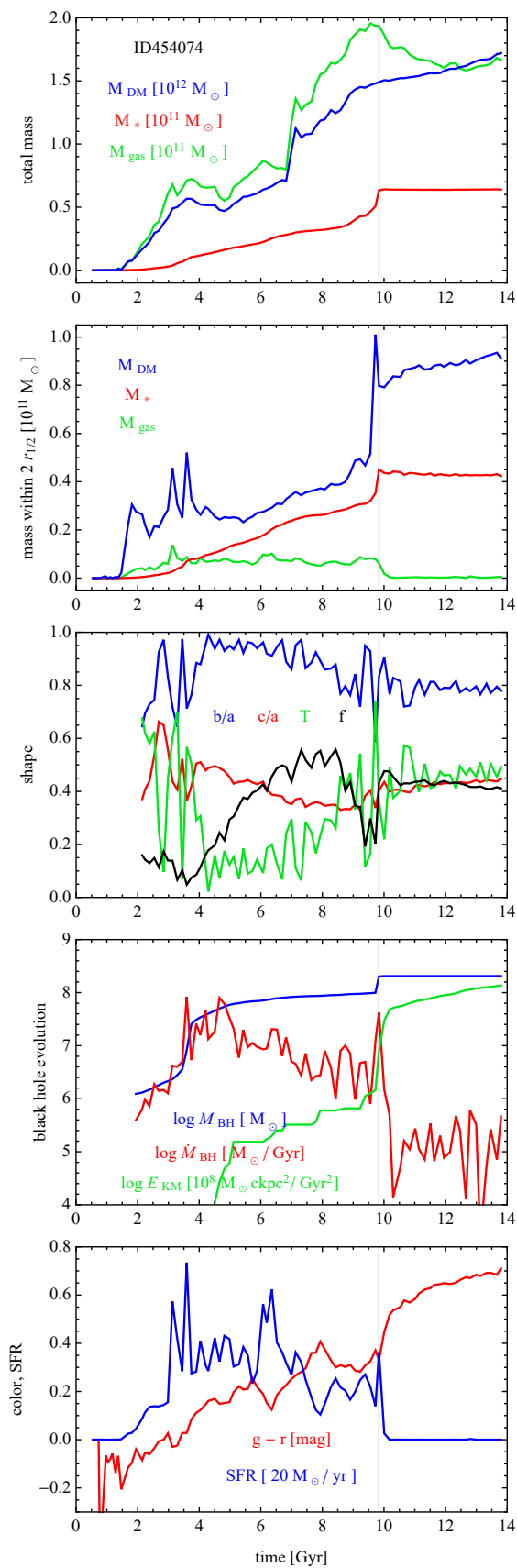}
\caption{Evolution of galaxy ID454074. Subsequent panels (from top to bottom) show the evolution of the
total mass in the various components, mass of the same components within $2 r_{1/2}$, shape and black hole
properties, as well as the color and SFR. The vertical gray line in each panel indicates the time of the merger.}
\label{evolution}
\end{figure}

We recently used the IllustrisTNG simulations in the 100 Mpc box (TNG100) to select and study a population of red
spirals \citep{Lokas2022b}. We found that only a surprisingly low fraction of 13\% of these objects could have formed
as a result of strong environmental effects, such as ram-pressure and tidal stripping of the gas in clusters. Instead, a
majority of them showed correlations between the timescales of the black hole growth and the gas loss, suggesting
that AGN feedback is responsible for quenching star formation in passive spirals. At the same time, the simulations
reproduced the observed trend of red spirals being more likely to possess bars. In some cases of barred red spirals, the
bar starts to form at the time when the gas starts to get depleted due to the AGN feedback in the kinetic mode. This
suggests that there may be a causal relation between the AGN feedback and the formation of the bars. Here, we use a
telltale example of a red spiral from this sample, albeit with a particular formation scenario, in order to determine
if  (at least in some cases) the quenching of galaxies by AGN feedback may indeed induce bar formation.

\begin{figure}
\centering
\includegraphics[width=7.1cm]{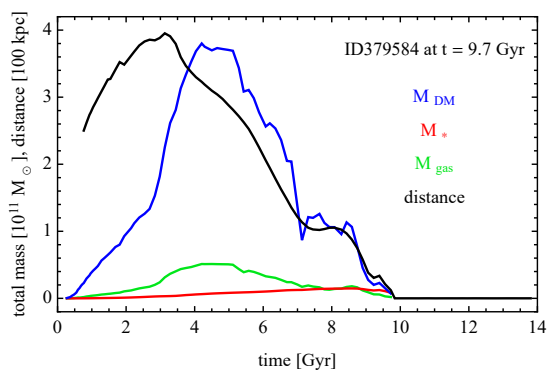}
\caption{Evolution of the satellite galaxy identified as ID379584 in the last snapshot before the merger. }
\label{evolutionsat}
\end{figure}

\section{The merger}

The galaxy studied here is identified as subhalo number ID454074 in the last snapshot of the Illustris TNG100
simulation, corresponding to the present time of $t = 13.8$ Gyr or redshift $z=0$. Data on this object were extracted
from the publicly available dataset originating from the highest resolution simulation performed in the 100 Mpc box of
the IllustrisTNG project (TNG100-1), as described by \citet{Nelson2019}. It is one of the red spirals studied
previously by \citet{Lokas2022b}, namely, a sample of disk galaxies with the color $g - r > 0.6$. The galaxy is unique
in terms of its formation scenario, as it formed in an almost equal-mass major merger.

The evolution of the main galaxy, which is the bigger of the two participating in the merger, followed along its main
progenitor branch, is illustrated in Fig.~\ref{evolution}. In the upper panel, we plot the total mass of the
object in different components: dark matter, $M_{\rm DM}$ (blue line), stars, $M_*$ (red), and gas, $M_{\rm gas}$
(green). For clarity, the units of the dark matter mass were chosen to be an order of magnitude larger than the values
for the gas and stars. The mass of the galaxy increases almost monotonically over time, especially for the dark matter
and stars.

The same properties for the satellite galaxy that eventually merges with the main object are shown in
Fig.~\ref{evolutionsat}. In this case, the mass of the most massive components, namely, the dark matter and gas, grows only
until $t = 4.2$ Gyr. The mass then starts to decrease due to tidal and ram-pressure stripping as the satellite starts
to approach the main galaxy. At this time, both galaxies have comparable total masses, with a mass ratio
of 1.4; while the maximum total mass of the satellite is $4.4 \times 10^{11}$ M$_\odot$, the total mass of the bigger
galaxy is then only $5.9 \times 10^{11}$ M$_\odot$. The relative distance, $d,$ of the two progenitors is shown with the
black line. The most spectacular drop in the dark matter mass of the satellite takes place at $t = 7$ Gyr, at the first
pericenter of $d = 100$ kpc.

The satellite then spirals into the bigger galaxy and merges with it at $t = 9.8$ Gyr, marked by the gray vertical line
in all panels of Fig.~\ref{evolution}. We define the merger time as corresponding to the simulation snapshot when the
satellite is no longer identified as a self-bound subhalo by the \textsc{subfind} algorithm \citep{Springel2001}. The
dark matter and gas are gradually lost by the satellite and absorbed by the main galaxy between $t = 4$ Gyr and the
time of the merger; whereas the stars are retained for a longer period because they are more tightly bound and they are only
swallowed by the bigger object at the merger time.

The second panel from the top of Fig.~\ref{evolution} provides a clearer explanation of what happens inside
the main body of the galaxy. Here, we plot the masses of the different components inside twice the stellar half-mass
radius, $2 r_{1/2}$. This radius grows from around 3 kpc at $t = 4$ Gyr to 6.26 kpc at the end of the simulation, with
the strongest increase at the time of the merger. The stellar mass component within $2 r_{1/2}$ (as the one that is the
most concentrated) behaves in a similar way as the total stellar mass, while the dark matter shows a strong increase
during and after the merger and the gas is almost completely removed.

\begin{figure}
\centering
\hspace{0.35cm}
\includegraphics[width=5.58cm]{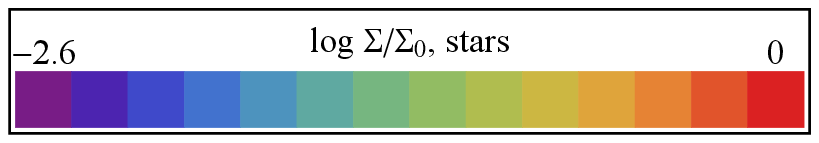}\\
\includegraphics[width=7cm]{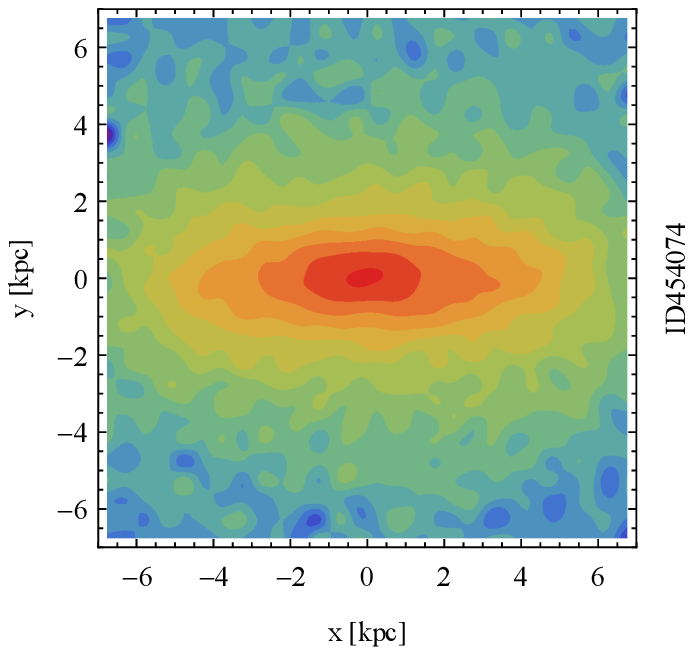}\\
\vspace{0.3cm}
\hspace{0.35cm}
\includegraphics[width=5.58cm]{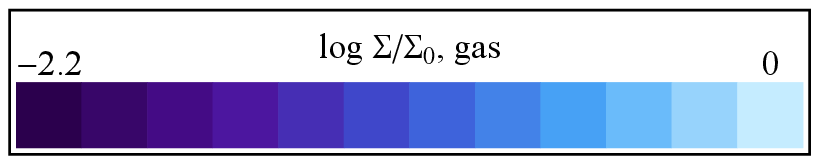}\\
\hspace{-0.21cm}
\includegraphics[width=7.187cm]{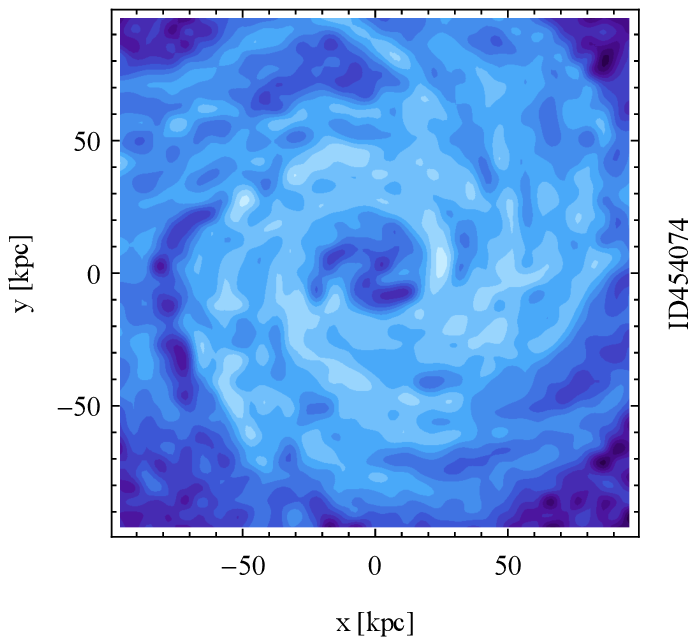}
\caption{Surface density distribution of the stellar component (upper panel) and the gas (lower panel) of
galaxy ID454074 in the face-on view at the present time. The surface density, $\Sigma,$ is normalized to the maximum
value, $\Sigma_0$, in each case and the contours are equally spaced in $\log \Sigma/\Sigma_0$.}
\label{surden}
\end{figure}

\section{Formation of the bar}

The third panel of Fig.~\ref{evolution} illustrates the evolution of the shape and kinematics of the main galaxy. The
shape is described using the shortest-to-longest and intermediate-to-longest axis ratios, $c/a$ and $b/a$, determined
from the mass tensor of stellar particles within $2 r_{1/2}$ \citep{Genel2015}, as well as the triaxiality parameter
$T = [1-(b/a)^2]/[1-(c/a)^2]$, shown with the red, blue and green lines, respectively. We can see that after the
initial period of strong variation, at around $t = 4$ Gyr, the shape parameters become stable and characteristic of oblate
spheroids or disks, with $b/a$ close to unity and $T < 1/3$. The parameters vary strongly during the merger as the disk
is destroyed and stabilize again afterwards, but this time with $b/a < 1$ and $T > 1/3$, which signifies the presence
of the triaxial shape.

\begin{figure}
\centering
\hspace{0.3cm}
\includegraphics[width=3.5cm]{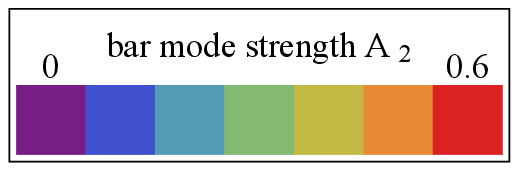}
\includegraphics[width=8.9cm]{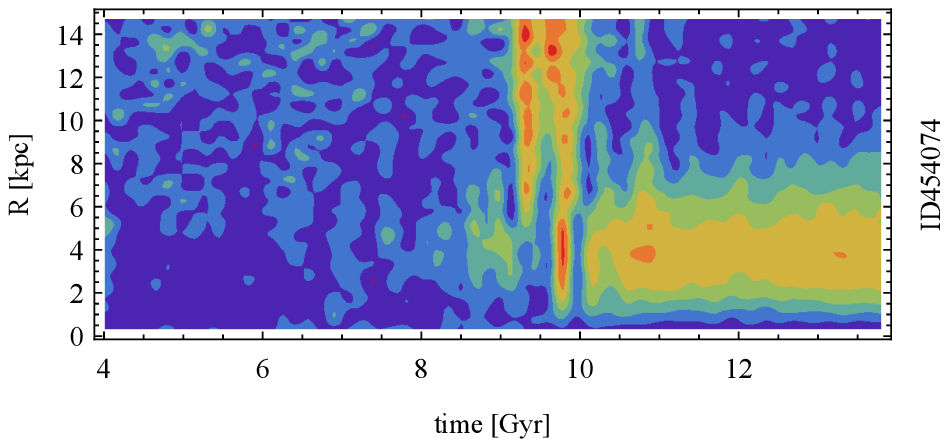}
\caption{Evolution of the profiles of the bar mode, $A_2 (R),$ of galaxy ID454074 over time.}
\label{a2modestime}
\end{figure}

The black line in the third panel of Fig.~\ref{evolution} shows the evolution of the rotation parameter, $f,$ defined as
the fractional mass of all stars with circularity parameter $\epsilon > 0.7$, where $\epsilon=J_z/J(E)$, $J_z$ is the
specific angular momentum of the star along the angular momentum of the galaxy, and $J(E)$ is the maximum
angular momentum of stellar particles for a given binding energy. The value of this parameter
grows steadily between $t = 4$ Gyr and the time of the merger reaching values of $f > 0.4,    $  which is typically adopted as
characteristic of rotationally supported disks of IllustrisTNG galaxies \citep{Joshi2020, Lokas2022a, Lokas2022b}. It
decreases significantly during the merger since the disk is strongly perturbed, but recovers at a level of $f > 0.4$
afterwards.

The triaxial shape detected using the axis ratios after the merger takes the form of a strong bar embedded in the disk that is fully visible in the final simulation output in the face-on projection of the stellar component (shown in the upper panel
of Fig.~\ref{surden}). The formation of the bar is even better described by the calculation of the profiles of the bar
mode $A_2$, that is the $m=2$ mode of the Fourier decomposition of the surface density distribution of stellar
particles projected along the short axis. The profiles were obtained from $A_m (R) = | \Sigma_j m_j \exp(i m \theta_j)
|/\Sigma_j m_j$, where $\theta_j$ is the azimuthal angle of the $j$th star, $m_j$ is its mass, and the sum goes up to
the number of particles in a given radial bin along the cylindrical radius, $R$.

The profiles were calculated for each simulation output in bins of $\Delta R = 0.5$ kpc and the results for the
dependence of the bar mode on both time and radius, $A_2 (R, t)$, are shown in the color-coded form in
Fig.~\ref{a2modestime}. The merger manifests itself clearly in this image as the strong elongation of the stellar
component at $t = 9.8$ Gyr and a stable bar forms soon afterwards, around $t = 10.5$ Gyr. We note that a weak bar
appears in the galaxy already around $t = 9$ Gyr, namely, before the merger; this is either  the result of the intrinsic
instability of the disk or tidal interaction with the same satellite. The bar is weak in comparison
with the one at the end of evolution, with the maximum of the bar mode profile, $A_{2, \rm{max}}$, on the order of 0.3;
while at the end, it reaches almost 0.5. The figure can also be used to estimate the length of the bar at different
times, as the radius where the value of $A_2$ drops to half the maximum. At the final simulation output, this length is
about 7 kpc, which agrees with the visual impression from the image in the upper panel of Fig.~\ref{surden}.

\section{The effect of AGN feedback}

It is interesting to ask why the bar forms so fast after the merger in this galaxy. In this section, we provide
evidence supporting the crucial role played by the AGN feedback in this process. As mentioned above, after the merger, the
main body of the galaxy (within $2 r_{1/2} = 12.5$ kpc) is almost completely deprived of gas. The mass of the gas
remaining in this region is on the order of $5 \times 10^8$ M$_\odot$, which comprises only 1\% of the mass of baryons
(gas and stars) there.

The most obvious candidate for the process responsible for such an efficient removal of the gas is the feedback from the AGN.
Each of the merging galaxies comes with its own supermassive black hole and they also merge. As a result, the black
hole mass $M_{\rm BH}$ of the main galaxy increases from $9.9 \times 10^7$ M$_\odot$ before the merger to more than
twice this value, namely, $2 \times 10^8$ M$_\odot$, afterwards. This is illustrated by the discontinuous increase of the values
shown with the blue line in the fourth panel of Fig.~\ref{evolution}. This increase causes the switch of AGN feedback
from the thermal to the kinetic mode and low accretion rate, $\dot{M}_{\rm BH}$, which indeed drops after a very short
peak during the merger, as shown by the red line in the same panel. The green line illustrates the
evolution of the cumulative amount of AGN feedback energy injected into the gas in the kinetic mode, $E_{\rm KM}$,
and we see that its increase during the merger is indeed sizable. On the other hand, the cumulative energy in the
thermal mode (not shown), although larger, mainly grew in an earlier phase and flattened around 6 Gyr.

This behavior is expected from the implementation of the AGN feedback adopted in the
IllustrisTNG simulations \citep{Weinberger2017, Weinberger2018}, where the kinetic mode starts to be important when the
black hole mass exceeds about $10^8$ M$_\odot$. The feedback energy was injected into the surrounding gas removing it
from the inner part of the galaxy. This scenario is confirmed by the face-on surface density of the gas, shown in the
lower panel of Fig.~\ref{surden}. Indeed, a gap in the gas distribution is visible in the central part, in spite of
the fact that the galaxy retains a large amount of gas overall, with a total mass of $1.7 \times 10^{11}$
M$_{\odot}$. We note that the gaseous disk is much larger than the stellar one, as indicated by the different scales of
the two images in Fig.~\ref{surden}.

The removal of the gas caused the star formation rate (SFR) in the inner region (within $2 r_{1/2}$) to drop to zero
after a very short increase taking place during the merger, as shown in the last panel of Fig.~\ref{evolution} (blue line). Due to
the lack of star formation, the galaxy quickly became red, with its color reaching $g - r > 0.6$ (red line); this is
above the threshold usually adopted to classify galaxies as red \citep{Nelson2018, Mahajan2020, Lokas2020b}. We note
that the colors of the IllustrisTNG galaxies (e.g., the $g$ and $r$ colors from the SDSS $ugriz$ rest-frame
bands) were estimated from all stars in a given object from the catalogs of synthetic stellar photometry whose calculations include the effects of dust obscuration \citep{Nelson2018}.
The removal of the gas caused by AGN feedback must have contributed significantly to the formation of the bar, since
low gas fractions in the inner regions of galaxies favor the formation of such a component
\citep{Athanassoula2013, Lokas2020a, Lokas2021}.

\section{Discussion}

Contrary to earlier beliefs, mergers of gas-rich disks are now recognized as one of the formation channels
of spiral (rather than only elliptical) galaxies, including those containing bars \citep{Athanassoula2016, Sparre2017,
Peschken2020}. While the presence of a significant amount of gas (also in the halo) has been identified as a necessary
condition to re-form a disk after the merger, the role of AGN feedback and, in particular, its relation to the
formation of the bar are not yet fully understood. However, \citet{Athanassoula2016} pointed out the necessity to
introduce a prescription for AGN feedback in their simulations of mergers to prevent the formation of a central
mass concentration that would prohibit the formation of the bar.

To shed more light on this issue, we analyzed the evolution of a spiral galaxy forming via a merger in the
Illustris TNG100 simulation. Due to the specific initial conditions of the merger, with both galaxies entering the
interaction with their own supermassive black holes, we could clearly witness the effect of the AGN feedback. Following the prescription for the feedback applied in the simulations, immediately after the two black holes
merge, the galaxy switches to the kinetic feedback mode which removes almost all the gas from its inner region. Soon
after the bar forms, taking advantage of the very low gas fraction in the center.

This process, which involves the causal relation between the AGN feedback and the formation of the bar, may in fact be quite
general and may also occur in other barred galaxies that have not experienced a major merger. For example, among the red
spirals studied previously \citep{Lokas2022b}, we identified many cases of AGN switching to kinetic feedback and a bar
starting to form, however, usually such events took place concurrently within some period of time and their causal relation
was more difficult to discern.

The scenario for the AGN-bar relation proposed here agrees with the correlations (or lack thereof) previously reported in
the literature. It explains the higher frequency of bars among the red spirals which were demonstrated to be mostly
quenched by AGN feedback \citep{Lokas2022b}. It also agrees with the lack of correlation between the strength of the
AGN emission and the presence of the bar \citep{Cisternas2015, Galloway2015, Cheung2015}. We find that it is not the bars that feed
the AGNs but, rather, AGNs that favor the formation of the bars. By the time the bar forms, the AGN is already past the
strong thermal mode characteristic of the quasar phase and, indeed, we found that red spirals with bars on average have
more massive black holes and lower accretion rates characteristic of the kinetic mode \citep{Lokas2022b}.
\citet{Rosas2020} also found that the black hole masses and their energy release in the kinetic mode are
higher in their sample of massive strongly barred galaxies in TNG100, in comparison to those without bars.

The relation between the presence of the bar and quenching is also clarified by this picture. We reproduced the central
hole in the gas distribution identified in some gas-rich massive spirals \citep{Newnham2020}, which had been
interpreted as supportive of the bar quenching mechanism, in which bars form and remove the gas from the center.
However, it was shown by \citet{Khoperskov2018} that bars induce turbulence in the gas and stabilize it
against collapse, thereby increasing its dispersion rather than expelling it out of the galaxy. In the scenario
proposed here, the gas is removed from the galaxy by AGN feedback and it is only later that the bar forms, due to the
low gas fraction.

These conclusions are certainly subject to the assumption that the AGN feedback implementation used in IllustrisTNG is
realistic. It remains to be seen whether other studies, using different black hole feedback models, also find similar
correlations. For example, \citet{Bonoli2016} and \citet{Spinoso2017} found that the bar is stronger in their Eris
simulations of the Milky Way if a massive black hole is present, although in their case the feedback is
rather weak and affects only the inner 1-2 kpc. An opposing conclusion, however, was recently reached by
\citet{Irodotou2022} using \textsc{auriga} simulations of Milky Way-like objects, which may be due to different merging
histories of the galaxies, different hydrodynamical schemes, or details of the subgrid physics. The
processes of AGN feedback, bar formation, and quenching certainly merit further study to clarify their
interplay.

\begin{acknowledgements}
I am grateful to the IllustrisTNG team for making their simulations publicly available.
\end{acknowledgements}

\end{document}